\documentstyle[preprint,tighten,eqsecnum,aps]{revtex}
\begin{document}

\title{Effective dynamics of self--gravitating extended objects}

\author {S.Ansoldi\footnote{E-mail address:
ansoldi@vstst0.ts.infn.it}}
\address{ Dipartimento di Fisica Teorica 
dell'Universit\`a,\\
Strada Costiera 11, 34014-Trieste, Italy,}

\author{ A.Aurilia\footnote{E-mail address:
aaurilia@csupomona.edu}}
\address{ Department of Physics, California State 
Polytechnic 
University\\
Pomona, CA 91768, USA,}

\author{ R.Balbinot\footnote{E-mail address:
balbinot@bologna.infn.it}}
\address{Dipartimento di Fisica dell'Universit\`a,\\
Istituto Nazionale di Fisica Nucleare, Sezione di 
Bologna,\\
via Irnerio 46, 40126-Bologna, Italy,}

\author{ E.Spallucci\footnote{E-mail address:
spallucci@vstst0.ts.infn.it}}
\address{Dipartimento di Fisica Teorica dell'Universit\`a,\\
Istituto Nazionale di Fisica Nucleare, Sezione di 
Trieste,\\
Strada Costiera 11, 34014-Trieste, Italy,}
\maketitle        
\newpage
\begin{abstract}
We introduce an {\it effective Lagrangian} which describes the classical and 
semiclassical dynamics of spherically symmetric, self-gravitating objects that
may populate the Universe at large and small (Planck) scale. These include 
wormholes, black holes and inflationary bubbles. We speculate that such 
objects represent some possible modes of fluctuation in the primordial 
spacetime foam out of which our universe was born. Several results obtained 
by different methods are encompassed and reinterpreted by our effective 
approach. As an example, we discuss: i) the gravitational nucleation 
coefficient for a pair of Minkowski bubbles, and ii) the nucleation 
coefficient of an inflationary vacuum bubble in a Minkowski background
\end{abstract}
\newpage
\def\li{\Lambda_{ in}}
\def\lo{\Lambda_{ out}}
\def\si{\sigma_{ in}}
\def\so{\sigma_{ out}}
\def\bi{\beta_{ in}}
\def\bo{\beta_{ out}}
\def\g{l_{ P}}
\newcommand\beq{\begin{equation}}
        \newcommand\eeq{\end{equation}}
        \newcommand\en{energy momentum tensor}
        \newcommand\beqa{\begin{eqnarray}}
        \newcommand\eeqa{\end{eqnarray}}
\def\R{\dot R}
\def\p{P_{ R}}
\def\pe{P_{E}}

\newpage

 \section{Introduction} 
 According to some current ideas in cosmology, in particular the chaotic 
inflationary scenario\cite{linde}, the Universe may consist of infinitely 
many self--reproducing bubbles which are continuously nucleated quantum 
mechanically. Some of them expand and look like a Friedmann universe, others 
may collapse to form black holes, and some may be connected by wormholes. 
In principle, this {\it dynamical} network may exist at any 
scale of distance, or energy; at the Planck scale of energy, it is 
expected to arise from gravitational fluctuations which induce a 
foam--like structure over the spacetime manifold. Some work on the 
dynamics of the constituents of this network has already
appeared in the literature\cite{red,visser,koz,berez,viss2},
but the results so far obtained are not always consistent and seem 
to be model dependent. As a matter of fact, the very notion of spacetime 
foam, introduced by Wheeler nearly forty years ago\cite{wheeler}, has 
periodically come under close scrutiny in the intervening years. The point of 
fact is that a complete quantum theory of the above processes is still beyond 
our reach, and one must come up with some viable alternative.

The aim of this communication is to suggest an {\it effective} approach to the 
dynamics of spherically symmetric, self--gravitating objects that 
may arise, evolve and die in the spacetime foam. Among these 
``~objects~'', somehow, there is the universe in which we live, and 
therefore the study of such fluctuations hardly needs a 
justification. In order to give some analytical substance to this qualitative 
picture, we envisage the spacetime foam as an ensemble of vacuum bubbles, or 
{\it cells} of spacetime, each characterized by its own geometric phase
and vacuum energy density. In principle each cell may behave as a black hole, 
a wormhole, an inflationary bubble, etc., depending on the matching 
conditions on the neighboring cells \cite{smail}. Thus, a useful starting 
point for our present discussion, is the general matching equation between 
the internaland external metrics of a self-gravitating, 
spherically symmetric bubble \cite{noi}
\begin{equation}
\sigma_{in}R\,\sqrt{1-{\Lambda_{in}\over 3}R^2+\dot R^2}-
\sigma_{out}R\,\sqrt{1-{2MG_N\over R}-{\Lambda_{out}\over 3}R^2+\dot 
R^2}=
4\pi\rho\,G_N R^2\ .
\label{2.1.1}
\end{equation}
In the above equation, $\Lambda_{ in,\, out}=8\pi 
G_N\,\epsilon_{in,\, out}$, 
are the cosmological constants representing
the internal(in) and external(out) vacuum pressures;  $\rho$ is the constant
surface tension, and $\sigma_{ in}(\sigma_{out})=\pm 1$ depending on 
whether the radius of the bubble
increases, or decreases, along the outward normal direction to the 
2-brane surface embedded in the interior (exterior) metric. 
Here we have assumed, for the sake of simplicity, the membrane equation of 
state $\rho=-p=constant$. However, we should emphasize that other 
equations of state, such as $p=0$ for dust, can be easily accomodated 
in our formalism. Following a suggestion first made in Ref.\cite{koz}, we 
now propose to interpret Eq.(\ref{2.1.1}) as the ``~Hamiltonian constraint~'', 
$\displaystyle{{\cal K}=0}$, for a system described by 
the {\it effective hamiltonian}, or super-hamiltonian,

\begin{equation} 
{\cal K}\equiv 4\pi\rho\,R^2-
\sigma_{in}{R\over G_N}\,\sqrt{1-{\Lambda_{in}\over 3}R^2+\dot R^2}+
\sigma_{out}{R\over G_N}\,
\sqrt{1-{2MG_N\over R}-{\Lambda_{out}\over 3}R^2+\dot R^2}
\ .
\end{equation}
Then, we can use the Hamilton equation, 
$\displaystyle{d{\cal K}=\dot R\, dP_R}$, 
to deduce the momentum $P_R$ which is canonically conjugated to 
$\dot R$, and the corresponding lagrangian
\begin{eqnarray}
P_R&=&\int {\partial {\cal K}\over \partial\dot R}{d\dot R\over \dot 
R}\ ,
\label{2.1.2}\\
L^{\rm eff}&=& P_R\dot R- {\cal K}\label{2.1.3}\ .
\end{eqnarray}
Thus, from the condition (\ref{2.1.1}), we find
\begin{equation}
P_R=
{R\over G_N}\ln\Bigg\vert
\left({\beta_{out}\over\beta_{ in}}\right)^{1/2}
{\dot R+\sigma_{ in}\sqrt{\dot R^2+\beta_{ in}}\over\dot R+
\sigma_{out}\sqrt{\dot R^2+\beta_{out}}}\Bigg\vert
\label{2.1.4}
\end{equation}
where 
$\displaystyle{\beta_{ in}\equiv 1-(\Lambda_{ in}/3)R^2}$ and 
$\displaystyle{\beta_{out}\equiv  1-(\Lambda_{out}/3)R^2-2MG_N/R}$.

The inverse Legendre transform (\ref{2.1.3}) leads to the 
{\it effective lagrangian} we are looking for
\begin{eqnarray}
L^{\rm eff}&=&{R\over G_N}\left(\,\sigma_{ in}\sqrt{\dot R^2+\beta_{ 
in}}-
\sigma_{out}\sqrt{\dot R^2+\beta_{out}}
-4\pi\rho\, G_N R\right)+\nonumber\\
&-&{R\dot R\over G_N}\left[
\sigma_{ in}\sinh^{-1}\left({\dot R\over\sqrt{\beta_{ in}}}\right)-
\sigma_{out}\sinh^{-1}\left({\dot R\over\sqrt{\beta_{out}}}\right)
\right]\ .
\label{2.1.5}
\end{eqnarray}
From here, we deduce the {\it proper time effective hamiltonian} in 
canonical form,
\begin{equation}
{\cal H}=4\pi\rho\, R^2-sign(\rho){R\over G_N}\left[\,\beta_{ 
in}+\beta_{out}-
2(\beta_{ in}\beta_{out})^{1/2}\cosh\left({G_N P_R\over R}
\right)\right]^{1/2}\ .
\label{2.1.6}
\end{equation}
At this point, we should mention that the above 
Lagrangian and Hamiltonian can also be deduced, in a more fundamental way, 
directly from the Einstein-Hilbert action  plus a boundary term,
under the assumption of spherical symmetry. This can be verified, for instance,
 by extending the derivation of ref.\cite{visser}
to our general case. The variational principle, in this case, leads 
precisely to the equations of motion derived from the Hamiltonian constraint 
 $\displaystyle{{\cal K}=0.}$ It is also worth observing that, by 
our reformulation, we have transformed the initial problem, i.e., 
the motion of a spherically symmetric self-gravitating membrane, into an 
equivalent, one dimensional, non--linear problem involving the dynamics of 
a {\it single degree of freedom}. In this ``~point--particle~'' 
interpretation of our effective lagrangian, one expects that 
``~pair production~'' takes place, and in section II we discuss an 
explicit example of this process. Finally, we should mention that the 
formulation given in ref.\cite{guth} 
is conceptually close to  ours, but is based on the use of
the Schwarzschild coordinate time as evolution parameter along the
bubble trajectory; the drawback of this choice of coordinate is that, even 
assuming a vanishing external pressure, it is impossible to Legendre 
transform the effective lagrangian and obtain the corresponding hamiltonian as
a function of the canonical pair $(R,\,P_R)$. 

Note that the hamiltonian (\ref{2.1.6}) involves a square root 
operation in analogy to the familiar expression of the 
energy of a relativistic point particle. 
In our case the coefficient of the square root depends on 
$sign(\rho)$ in order to be consistent with the {\it classical} 
equation of motion (\ref{2.1.1}). The opposite sign 
is classically meaningless. However, since we are dealing with a 
relativistic system, both positive and negative energies become physically 
relevant at the quantum level. {\it This leads us to a broader interpretation 
of spacetime foam as a ``~Dirac sea of extended objects~'', in which 
not only wormholes, but also black holes and vacuum bubbles are 
continuously created and destroyed as zero-point energy 
fluctuations in the gravitational quantum vacuum}. As a matter of 
fact, the effective lagrangian (\ref{2.1.5}), (or the effective hamiltonian
(\ref{2.1.6})), encodes the dynamics of a {\it four--parameter ($\rho$, 
$\Lambda_{in}$, $\Lambda_{out}$, $M$) family} of 
spherically symmetric, classical solutions of the self--gravitating
bubble equation of motion. Furthermore, the results (\ref{2.1.5})
and (\ref{2.1.6}) can be extended in a straightforward manner to an even 
larger family of solutions by endowing the internal geometry with a 
non--vanishing Schwarzschild mass term  $M_{\rm in}G_N/R$,
or the external geometry with an electric charge.

The spacetime foam models currently available in the literature
focus essentially on the Schwarzschild metric 
(see, however, ref.\cite{red}), and correspond to the sector $M>0$ 
of our family of classical solutions.
For instance, in the sub--sector $\rho>0$, $\Lambda_{in}>0$,
$\Lambda_{out}=0$, one finds the vacuum bubbles discussed
in ref. \cite{guth1}. In particular, we recall the type E 
trajectories with $M>M_{cr}$, listed in the same reference, because 
they give rise to baby--universes connected through wormholes to 
the parent universe. The characteristics of those trajectories are 
instrumental to our discussion in Section IV.

Other types of ``~foam--like~'' solutions belong to the subsector
$\Lambda_{in}=\Lambda_{out}=0$. They include the ``~surgical~''
Schwarzschild--Schwarzschild wormholes \cite{visser}, and the 
``~hollow~'' Minkowski--Schwarzschild wormholes \cite{koz},
\cite{ber2}, while the region $\rho<0$ of the same subsector 
contains the {\it traversable} wormholes, i.e. wormholes whose throats can be 
crossed by timelike observers. In this connection, note that in our membrane 
model a negative tension plays the same role as the negative energy density 
of the more conventional wormholes made out of ``~dust~'': it 
provides the ``~repulsive force~'' required to oppose the 
gravitational collapse of the throat \cite{thorne}. The explicit 
correspondence between negative energy density and surface tension 
is provided by the simple 
relation: $\displaystyle{\rho\leftrightarrow m/4\pi R^2}$.  
However, while the negative energy density of dusty wormholes is 
ascribed to some kind of {\it exotic matter} \cite{thorne}, or to 
gravitational vacuum polarization \cite{roman}, we suggest 
to interpret membranes with negative tension as boundary layers 
between different physical vacua as suggested, for instance, by the 
existence of normal and confining vacua in QCD. More about negative 
energy density later.

In section II  we show that in the flat spacetime limit, i.e. for
$G_N\rightarrow 0$, the resulting lagrangian can be used to compute
the false vacuum decay amplitude in ordinary quantum field theory.

In sections III and IV, as an explicit application of our general 
method, we shall study two examples of gravitational fluctuations in 
the sector $\Lambda_{in}=\Lambda_{out}=M=0$, which correspond to 
{\it vacuum bubbles}. Their discussion and comparison with the 
existing literature on the subject provide an excellent 
testing ground for the validity of our approach.

\section{ False vacuum decay}

Our effective approach is general enough to include
bubble dynamics in the absence of gravity .As a matter of fact, the limit 
$G_N\rightarrow 0$
provides a necessary consistency check of our method and represents a 
special case worth investigating in and by itself.

The correct limiting procedure requires to express the two cosmological
constants in terms of the corresponding vacuum energy densities
$\epsilon_{\rm in}$, $\epsilon_{\rm out}$:
$\displaystyle{\li=8\pi G_N \epsilon_{\rm in}}$,\
$\displaystyle{\lo=8\pi G_N \epsilon_{\rm out}}$. In a single
Minkowski domain $\si=\so=1$. Then, by expanding
$L^{\rm eff}$ up to the first order in  $G_N/R^2$ we obtain,
\beq
L^{\rm eff}=-4\pi\rho R^2-{4\pi\over 3}\Delta\epsilon R^3\sqrt{1+\R^2}\ 
,\qquad
\Delta\epsilon\equiv\epsilon_{\rm in}-\epsilon_{\rm out}\ .
\label{2.2.1}
\eeq
This effective lagrangian represents the minisuperspace approximation of 
the gauge action
for the membrane
\beqa
I=&-&\rho\int d^3\xi \sqrt{-\gamma}-{e\over 3!}
\int d^4x\, J^{\mu\nu\rho}A_{\mu\nu\rho}
\nonumber\\
&-&{1\over 2\cdot 4!}\int d^4x F_{\mu\nu\rho\sigma}F^{\mu\nu\rho\sigma}
+{1\over 3!}\int d^4x\,
\partial_\mu \left(F^{\mu\nu\rho\sigma}A_{\nu\rho\sigma}\right)
\ ,\label{2.2.2}
\eeqa
where $\gamma$ is the determinant of the induced metric 
on the 2-brane world tube $x^\mu=x^\mu(\xi^a)$, $a=0,1,2$, and
$J^{\mu\nu\rho}(x)$ is the 2-brane current\cite{noi}.
The last term in (\ref{2.2.2}) is a total divergence ensuring
 that variations of the gauge potential $A_{\mu\nu\rho}$ 
will not produce unusual boundary terms due to the presence of the 2--brane
world tube. The corresponding hamiltonian is
\beqa
{\cal H}&=&4\pi\rho R^2 +{4\pi\over 3}\Delta\epsilon R^3\,
\sqrt{1-(3\p/4\pi\rho R^3)^2}
\ ,\label{2.2.3}\\
\p&=&-{4\pi\over 3}\Delta\epsilon R^3 {\R\over\sqrt{1+\R^2}}\ .
\eeqa
The classical trajectories describing true vacuum bubbles
are solutions of the hamiltonian constraint
$\displaystyle{{\cal H}= 0}$, stating that the total mass energy of a 
vacuum bubble is vanishing. From equation (\ref{2.2.3}) we see that
classical solutions, corresponding to bubbles with positive surface
tension, are allowed only if $\Delta\epsilon<0$, that is, 
the internal energy density (~of the true vacuum~) must be smaller
than the external energy density (~of the false vacuum~), the net amount
of energy released in the transition being converted into (positive)
kinetic energy of the bubble wall.
The semi-classical picture of the true vacuum domain nucleation
corresponds to a {\it classically forbidden} motion.
The classically unphysical tunneling trajectory
is a solutions of the {\it euclidean, } equation of motion
obtained via the Wick rotation $\displaystyle{\tau_E\equiv i\tau\ ; 
\pe\equiv iP}$

\beq
{\cal H}_E=0\ ,\quad
\Rightarrow \pe=4\pi\rho R^2 \left( 1-{R^2\over R_0^2}\right)^{1/2}\ ,
\qquad R_0={3\rho\over\Delta\epsilon}\ .
\label{2.2.4}
\eeq
Then, using the classical solution (\ref{2.2.4})
for $\pe$ in the WKB integral
for the calculation of the nucleation probability through tunnel effect, we 
obtain the standard result,
\beq
B=2\int_0^{R_0} dR \pe(R)=
8\pi\rho\int_0^{R_0}dR R^2 \left(1-{R^2\over R^2_0}\right)^{1/2}
={\pi^2\over 2}\rho R^3_0\ ,
\label{2.2.5}
\eeq
in agreement with the Coleman-De Luccia $B$ decay coefficient \cite{coleman}.

\section{ ``~Minkowski pair~'' creation}

The simplest example of gravitational vacuum fluctuation is given
by the spontaneous nucleation of a vanishing mass--energy shell in Minkowski
spacetime. Despite the triviality of both the internal and external geometries, 
the dynamics of such a process is nonetheless affected by the 
unavoidable ambiguities and technical problems of quantum gravity. 
To circumvent all these difficulties, one usually employs a semi--classical 
model obtained by gluing together two Minkowski metrics along the bubble 
trajectory .

In our approach, the steps are as follows. The signs of 
$\sigma_{ in}\ ,\sigma_{out}$ are fixed by the matching condition
\begin{equation}
\sigma_{ in}R\sqrt{1+\dot R^2}-\sigma_{out}R\sqrt{1+\dot R^2}=
4\pi\rho\, G_N R^2 .
\label{2.3.1}
\end{equation}
For a positive surface tension we have 
$\displaystyle{\sigma_{ in}=-\sigma_{out}=1}$, otherwise 
the two spacetime domains cannot be glued
together. The resulting geometry represents the limiting configuration
of vanishing cosmological constant and Schwarzschild mass discussed
in appendix D of ref.\cite{guth1}. This is a closed universe formed by two
compact spherical regions of flat spacetime.  
The shell equation of motion obtained by squaring (\ref{2.3.1}) is
\begin{equation}
\dot R^2=-1+4\pi^2\rho^2G_N^2 R^2\ ,
\label{2.3.2}
\end{equation}
and admits only bounce solutions irrespective of the sign of $\rho$. 
In our case
\begin{equation}
L^{\rm eff}=-4\pi\rho R^2+{2R\over G_N}\sqrt{1+\dot R^2}-
{2R\dot R\over G_N}\sinh^{-1}\dot R\ ,
\label{2.3.3}
\end{equation}
\begin{equation}
P_R={\partial L^{\rm eff}\over \partial\dot R}=
-{2R\over G_N}\sinh^{-1}\dot R\ ,
\label{2.3.4}
\end{equation}

\begin{equation}
{\cal H}=4\pi\rho R^2 -{2R\over G_N}\cosh\left({G_N P_R\over 
2R}\right)\ .
\label{2.3.6}
\end{equation}
Note that, in order to recover Eq.(\ref{2.3.6}) from the general formula 
(\ref{2.1.6}), one must carefully assign the phases of the complexified 
functions $\sqrt{\beta_{in}}$, $\sqrt{\beta_{out}}$ in the analytically
extended Schwarzschild manifold, and only then one may consider the  
limit of vanishing Schwarzschild mass. 
This is because the effective hamiltonian
(\ref{2.1.6}) is actually a complex function when considered on the
maximally analytic extension of the underlying spacetime manifold. 
Note, also, that the kinetic term, obtained by expanding ``~$\cosh$~'' up 
to second order, is negative. Therefore, ${\cal H}=0$ is the classical 
equation of motion of a (positive kinetic energy) particle in the 
{\it reversed} potential $V(R)\sim -4\pi\rho R^3+2 R^2/G_N$. This potential 
constitutes an effective barrier for the ``~particle~'' motion and 
explains the absence of a discrete spectrum of stationary quantum 
states. The classical dynamics of
this type of domain wall has been discussed in ref.\cite{ipser} with
special emphasis on the repulsive character of the 
resulting gravitational field. In our formulation, this repulsive 
effect is plainly exhibited by the potential above.   
We propose to associate the expected quantum mechanical 
``~leakage~'' through the potential barrier with the process of {\it 
``~universe creation~'' by quantum tunneling from nothing.} 
In order to give substance to this interpretation, one needs to 
compute the corresponding transmission coefficient, and, in order to 
perform this calculation, we rotate 
the dynamical quantities to imaginary time:
\begin{equation}
P_E={2R\over G_N}\cos^{-1}\left({R\over R_G}\right)\ 
,\qquad\hbox{where}\qquad
R_G={1\over 2\pi \rho\, G_N}\ .
\label{2.3.7}
\end{equation}
This gives, for the nucleation coefficient

\begin{equation}
B={4\over G_N}\int_0^{R_G}dR\,R\,\cos^{-1}\left({R\over R_G}\right)=
{1\over 8\pi\rho^2 G_N^3}\ ,\label{2.3.8}
\end{equation}
in agreement with the bounce calculation of ref.\cite{ber2}.

In the case of negative surface tension the calculation proceeds 
along the same steps outlined above, but with 
$\sigma_{ in}=-1$, $\sigma_{out}=+1$. The resulting geometry is the limiting 
case of a ``~surgically~'' constructed Schwarzschild wormhole, when the 
mass is sent to zero \cite{visser}. However, any constant time section of the 
resulting spacetime has infinite volume. Thus, this second type of 
``Minkowski pair'' does not correspond to 
a compact object and cannot be nucleated quantum mechanically from "nothing".

\section{Inflationary bubble nucleation amplitude} 

The next case study is somewhat more subtle and we discuss it to 
illustrate the applicability of our effective method. To the extent 
that this system may arise as a fluctuation of the gravitational 
vacuum, we interpret it as a possible constituent of the Planckian 
spacetime foam even though it does not correspond to any class of 
wormholes. What we have in mind is the nucleation of a 
false vacuum de Sitter bubble in a Minkowski background which 
represents the key mechanism proposed in ref.\cite{guth1}, and 
expanded in ref.\cite{guth}, to generate quantum mechanically an 
inflationary domain. From our 
vantage point, the difficulty of that proposal is the presence of
a virtual black hole, decaying through Hawking radiation,
 as an intermediate state between the initial Minkowski
state and the final Minkowski$+$de Sitter state. This intermediate 
state involves the still obscure issue of the final stage
of black hole evaporation. Interestingly enough, it is possible to 
bypass this difficulty by choosing a {\it negative surface tension 
and vanishing total mass  energy} for the original quantum 
fluctuation that triggers the process in 
the first place. Physically, this assumption of negative surface 
tension may be justified by a simple analogy with the multiphase 
vacuum of QCD. If the unified field theory undergoing primordial 
phase transitions is of the Ginzburg-Landau type, then, for some 
choice of the coupling constants, bags can form around test charges 
with positive volume and {\it negative surface energy} 
\cite{hosek}. The sign of the surface tension follows from the 
negative condensation energy. The vacuum state for such a model
behaves as a type II superconductor with maximal 
boundary surface between the normal (~non--confining~) phase and
the ordered (~confining~) phase \cite{memorial}.

With the above choice of surface tension and vanishing total mass--
energy, the initial and final states are degenerate in energy and a 
spontaneous transition between them is allowed without an 
intermediate blackhole state.

Some preliminary remarks on the classical
dynamics of a de Sitter bubble will be helpful in order to clarify 
our final result. The classical equation of motion for the bubble trajectory is
\begin{eqnarray}
\dot R^2&=&-1+{R^2\over R_B^2}\ ,\label{2.4.4}\\
 R_B&=&{8\pi|\rho|\, G_N\over 16\pi^2\rho^2 G_N^2+1/ H^2 }
={R_0\over 1+R_0^2/4H^2}\ ,
\label{raggio}
\end{eqnarray}
where $H$ is the radius of the de Sitter cosmological event horizon, and 
$R_0=3|\rho|/\epsilon_{in}$ is the nucleation radius in the absence of 
gravity. Equation (\ref{2.4.4}) admits {\it only} a bounce-solution which 
represents  an infinite domain of de Sitter vacuum
collapsing down to a nonvanishing
minimum radius $R_B<R_0$, and then re-expanding indefinitely. 
Note also that $R_B\le H$, and that the equality holds only if the surface
 tension and the internal vacuum energy density are tuned so that
$\epsilon_{in}=6\pi \rho^2 G_N$. 
 
Having assumed a negative surface tension, we retrace our steps as in the 
previous case study: the matching condition now is
\begin{equation}
\sigma_{ in}R\,\sqrt{1+\dot R^2-R^2/ H^2}-
\sigma_{out}R\,\sqrt{1+\dot 
R^2}=
4\pi\rho\, G_N R^2
\ ,\qquad  H^2\equiv 3/\Lambda_{ in}\ ,
\label{2.4.1}
\end{equation}
and gives us
\begin{eqnarray}
\sigma_{out}&=&+1\ ,\label{2.4.2}\\
\sigma_{ in}&=&-\hbox{ sign}
\left(16\pi^2\rho^2G_N^2-{1\over H^2}\right)=-\hbox{ sign}(R_0-2H)\ 
.
\label{2.4.3}
\end{eqnarray}
Equations (\ref{2.4.2}) and (\ref{2.4.3}) show that, while the sign 
of $\sigma_{out}$
is fixed to $+1$ along the bubble trajectory, the sign of $\sigma_{ in}$ 
depends on the relative size of $R_0$ and $2H$. 
The two nucleation modes actually correspond to two different
Penrose diagrams, but we shall omit their discussion in this brief 
communication.

Presently, for reasons of clarity and conciseness, we choose to 
discuss the case $\displaystyle{R_0<2H\rightarrow \sigma_{ in}=+1}$. 
The corresponding semi-classical solution, which describes the nucleation of an 
expanding de Sitter bubble, is obtained by matching the expanding half of
the classical bounce to a quantum tunneling solution. Then, the classical
turning point acquires the meaning of nucleation radius. The 
corresponding lagrangian and hamiltonian are,

\begin{eqnarray}
L^{\rm eff}&=&{R\over G_N}\left(\sqrt{\beta_{ in}+\dot 
R^2}-
\sqrt{1+\dot R^2}-4\pi\rho\, G_N  R\right)+
\nonumber\\
&-&{R\dot R\over G_N}\left[
\sinh^{-1}\left({\dot R\over\sqrt{\beta_{ in}}}\right)-
\sinh^{-1}\dot R^2\right]\ ,
\label{2.4.5}\\
P_R&=&-{R\over G_N}\left[\sinh^{-1}\left({\dot R\over
\sqrt{\beta_{ in}}}\right)-\sinh^{-1}\dot R\right]
\label{2.4.6}\\
{\cal H}&=&4\pi\rho R^2+
{R\over G_N}\left[\beta_{ in} +1 +2
\sqrt{\beta_{ in}}\cosh\left({G_N P_R\over R}\right)\right]^{1/2}\ .
\label{2.4.7}
\end{eqnarray}
Note that we have fixed the phase of $\sqrt{\beta_{in}}$ and 
$\sqrt{\beta_{out}}$ by the condition that the above hamiltonian 
coincides with the hamiltonian (\ref{2.3.6}) in the limit of vanishing internal
energy density. The classical equation ${\cal H}=0$, gives
\begin{equation}
P_R=
{R\over G_N}\cosh^{-1}\left[{(4\pi\rho\, G_N R)^2-1-\beta_{ in}\over 
2\sqrt{\beta_{ in}}}\right]\ .
\label{2.4.8}
\end{equation}
This enables us to evaluate the tunneling amplitude
\begin{equation}
B=2\int_0^{R_B}dR |P_E(R)|
=-{1\over G_N}\int_0^{R_B}dR R^2{d\over dR}
\cos^{-1}\left[{(4\pi\rho\, G_N R)^2-1-\beta_{ in}\over 
2\sqrt{\beta_{ in}}}\right]\ ,
\label{2.4.9}
\end{equation}
and an explicit calculation yields
\begin{eqnarray}
B&=&4\pi|\rho|\int_0^{R_B} dR R^2\left(1-{R^2\over H^2}\right)^{-
1}\left(1
-{R^2\over R_B^2}\right)^{-1/2}\left({R^2\over  4\pi|\rho|\, G_N R_B 
H^2}-1\right)
\nonumber\\
&=&{\pi^2|\rho|\over 2} R_B^2 R_0\ .
\label{2.4.11}
\end{eqnarray}

Our last comment concern the physical interpretation of this result. 
The model discussed above describes the quantum birth of an 
inflationary bubble in a Minkowski background. In connection with this process, 
we find that there is a lingering ambiguity in the published 
literature. It is indeed interesting, and perhaps somewhat puzzling, 
that the initial radius and the nucleation rate in this case are the 
same as for the false vacuum decay, that is, the nucleation of a 
Minkowski bubble in a De Sitter background, originally discussed by 
Coleman and De Luccia \cite{coleman}. With hindsight, this 
coincidence is hardly surprising since the two cases 
appear to be completely symmetrical due to the fact that the 
euclidean trajectory interpolating between the two vacuum states is 
the same in both cases. However, there {\it is} a difference, even 
at the classical level, and it 
lies in the global structure of the spacetime manifold in the two 
cases. The point is that, for an inflationary bubble in a Minkowski 
background, at a given instant, say the nucleation  Minkowski time, 
all the points in the interval $0\le r\le R_B$ suddenly undergo a 
{\it phase transition}
from the Minkowski to the de Sitter geometry. Then, the new vacuum 
domain, driven by the negative pressure of the false vacuum, 
expands exponentially, eventually filling up the whole spacetime.  
In contrast, a true vacuum bubble, no matter how large, 
will never fill up the whole de Sitter manifold. As a matter of 
fact, in this difference lies the problem of the ``~graceful exit~'' 
from the inflationary stage \cite{gw}.

\end{document}